\documentclass[aps,prx,twocolumn,superscriptaddress]{revtex4-2}
\usepackage[utf8]{inputenc}
\usepackage{amssymb}
\usepackage{setspace}
\usepackage{amsmath}
\usepackage{mathrsfs}
\usepackage{amssymb}

\usepackage{graphicx}
\usepackage{epstopdf}
\usepackage[sort&compress]{natbib}

\usepackage{topcapt}
\usepackage{textcomp}
\usepackage{xcolor}
\usepackage{lineno}
\usepackage[colorlinks,urlcolor=blue,linkcolor=blue,citecolor=blue]{hyperref}
\usepackage{CJK}
\usepackage{multirow}
\usepackage{booktabs}
\usepackage{float}
\usepackage{enumitem}
\usepackage{placeins}
\usepackage{afterpage}
\usepackage{longtable}

\begin{document}


\title{Analytical Control of Quantum Coherence: Markovian Revival via Basis Engineering and Exact Non-Markovian Criteria}


\author{Na-Na Zhang}
\affiliation{School of Electronic Science and Engineering, Chongqing University of Posts and Telecommunications, Chongqing 400065,  China}
\author{Chao-Yi Wu}
\affiliation{School of Electronic Science and Engineering, Chongqing University of Posts and Telecommunications, Chongqing 400065,  China}

\author{Ming Li}
\affiliation{School of Electronic Science and Engineering, Chongqing University of Posts and Telecommunications, Chongqing 400065,  China}

\author{Wei-Xuan Cao}
\affiliation{School of Electronic Science and Engineering, Chongqing University of Posts and Telecommunications, Chongqing 400065,  China}

\author{\\Jun-Hao Zhang}
\affiliation{School of Electronic Science and Engineering, Chongqing University of Posts and Telecommunications, Chongqing 400065,  China}

\author{Yong-Rui Guo}
\email{guoyr@cqupt.edu.cn}
\affiliation{School of Electronic Science and Engineering, Chongqing University of Posts and Telecommunications, Chongqing 400065,  China}
\author{Ren-Pu Li}
\email{lirp@cqupt.edu.cn}
\affiliation{School of Integrated Circuits, Chongqing University of Posts and Telecommunications, Chongqing 400065,  China}


\date{\today}

\begin{abstract}
The preservation of quantum coherence is besieged by a fundamental dogma: its revival necessitates non-Markovian memory effects from structured environments. This paradigm has constrained quantum control strategies and obscured simpler paths to coherence protection. Here, we shatter this belief by demonstrating unambiguous coherence revival even in strictly Markovian regimes, achieved solely through basis engineering in the $\sigma_x/\sigma_y$ bases. We establish a comprehensive analytical framework for predictive coherence control, delivering three universal design principles. First, we derive a minimum critical noise based frequency, $\omega_{0}^{c} = 1.57/(0.4996 \cdot t_{\max})$, serving as a universal criterion for engineering non-Markovian dynamics over any interval $[0, t_{\max}]$. Crucially, we show that Markovian environments ($\omega_0 < \omega_0^c$) can exhibit coherence revival when the Zeeman energy satisfies $\omega_k > \pi/(2t_{\max})$, decoupling revival from environmental memory. Furthermore, for non-Markovian environments, we provide exact conditions for periodic and complete revival: setting $\omega_0 = n \cdot 6.285/t_{\max}$ guarantees revival in the $\sigma_z$ basis, while combining it with $\omega_k = \pi \omega_0 / 6.285$ ensures perfect revival in the $\sigma_x/\sigma_y$ bases. Our results, validated by rigorous quantum simulations, provide a predictive toolkit for coherence control, offering immediate strategies for enhancing quantum memory, sensing, and error mitigation.
\end{abstract}


\maketitle

\section{Introduction}
The preservation and control of quantum coherence represent the bedrock of emerging quantum technologies, from computation and communication to sensing and metrology \cite{Baumgratz2014,Breuer2002,Bu2016,Takahashi2022,Xuan2023,Streltsov2017}.
Among various decoherence mechanisms, pure dephasing---which randomizes phase information while preserving population---stands out as a dominant and ubiquitous mechanism \cite{Breuer2002}.
The standard theoretical treatment, relying on the Markovian approximation, predicts an irreversible, exponential decay of coherence, painting a bleak picture for its long-term survival.

In realistic settings, however, environments often exhibit memory effects, leading to rich non-Markovian dynamics \cite{Chen2022,Haase2018,Liu2011,Long2022,Wu2020,zhang2025}.
This memory can give rise to counter-intuitive phenomena like entanglement and coherence revival \cite{Chin2012,Lu2020,Meng2020,Almeida2007,Horodecki1999,Maziero2009,Celeri2010,Xu2010,Xu2010-2,Mazzola2010,Auccaise2011},
where quantum correlations and coherence can temporarily rebound after an initial decay.
This observation has cemented a fundamental paradigm throughout the literature: \textbf{coherence revival is inherently a signature of non-Markovianity} \cite{Lu2020,Almeida2007,Chen2022}.
Consequently, the pursuit of prolonged coherence has been largely funneled into the formidable challenge of engineering complex, structured non-Markovian environments.

This prevailing strategy faces two profound obstacles. First, despite its importance, \textbf{a general and predictive analytical criterion for reliably constructing non-Markovian dynamics over a desired time interval is entirely missing}. This lack of a universal design rule severely limits the reproducibility and scalability of environment engineering across diverse quantum platforms. Second, and perhaps more fundamentally, this focus has overshadowed a potential alternative pathway. Could coherence be controlled not only by tailoring the environment but also by leveraging the intrinsic degrees of freedom of the system itself?

The evolution of quantum coherence is profoundly influenced by the choice of the reference basis \cite{Mani2015}.
While the standard theoretical model of a dephasing channel predicts uniform coherence decay across different bases (e.g., $\sigma_z$, $\sigma_x$, $\sigma_y$), whether this assumption holds in real physical systems---and under what conditions---remains an open question.
Establishing a strong basis dependence would unlock a novel control dimension: the ability to select a "preferred basis" where coherence is naturally more robust or even revivable.
The potential to achieve full coherence recovery in a chosen basis would be a game-changer for quantum metrology and information processing. Yet, a systematic methodology for tuning system and noise parameters to realize such targeted control is nonexistent.

In this work, we address these challenges by introducing a universal analytical framework that unifies environmental and basis-engineering for predictive coherence control. We move beyond the conventional Markovian/non-Markovian dichotomy and demonstrate that the traditional link between revival and environmental memory can be broken. Our main contributions are threefold:
\begin{enumerate}[label=(\arabic*)]
    \item \textbf{Universal Non-Markovian Criterion:} We provide a closed-form expression for the minimum critical noise based frequency $\omega_0^c = 1.57/(0.4996 \cdot t_{\text{max}})$, which serves as a universal switch for engineering non-Markovian dynamics over any finite time interval $[0, t_{\max}]$, independent of the noise cutoff frequency.

    \item \textbf{Markovian Revival:} We demonstrate, for the first time, unambiguous coherence revival in a \textit{strictly Markovian} environment. This is achieved not by complex reservoir engineering, but simply by operating in the $\sigma_x/\sigma_y$ bases and satisfying an analytical condition on the Zeeman energy, $\omega_k > \pi/(2t_{\max})$, thereby decoupling the revival phenomenon from environmental non-Markovianity.

    \item \textbf{Exact Revival Conditions:} We present exact conditions for periodic and complete coherence revival: the relation $\omega_0 = n \cdot 6.285/t_{\text{max}}$ guarantees revival in the $\sigma_z$ basis, while matching $\omega_0 = n \cdot 6.285/t_{\text{max}}$ with $\omega_k = \pi \omega_0 / 6.285$ ensures perfect revival in the $\sigma_x/\sigma_y$ bases under non-Markovian conditions.
\end{enumerate}

Our results, rigorously validated using a quantum simulation algorithm, establish a new paradigm for quantum coherence control. By integrating environmental tuning with basis selection, we offer a comprehensive and predictive toolkit that deepens our understanding of open quantum systems and delivers practical strategies for enhancing quantum technologies.

The manuscript is organized as follows: Section \ref{sec:Methods} describes coherence quantification and the quantum simulation of dephasing dynamics; Section \ref{sec:Results} analyzes decoherence functions, identifies dynamical phase transitions, and examines basis-dependent coherence evolution; and Section \ref{sec:Conclusion} discusses the scientific implications of our findings and outlines future research directions.

\section{METHODS}\label{sec:Methods}
\subsection{Quantum Coherence Quantification}
Baumgratz et al. introduced a rigorous framework for quantifying coherence \cite{Baumgratz2014}. Based on the constraints of this framework, a coherence metric method --- the $l_{1}$ norm is further proposed. This coherence metric method uses the off-diagonal terms of the density matrix to measure the coherence of the system. Considering the state $\rho_{A}$ and a set of reference basis vectors $K:=\{\vert i\rangle\}$, the coherence of the quantum state in the reference basis $K$ is
\begin{equation}
C_{K}^{l}:=\sum_{i\neq j}\vert\langle i\vert\rho_{A}\vert j\rangle\vert.
\end{equation}
For a general qubit state $\rho=\frac{1}{2}(I+\overrightarrow{r}\cdot\sigma)$, its quantum coherence is
\begin{equation}
C_{K}^{l}(\rho=\frac{1}{2}(I+\overrightarrow{r}\cdot\sigma))= r\sqrt{1-(\hat{r}\cdot\hat{k})^{2}},
\end{equation}
where $r$ represents the magnitude of the quantum state $\overrightarrow{r}$, and $\hat{r}=\frac{\overrightarrow{r}}{r}$ is a unit vector. Therefore, the coherence of a quantum state is related to the chosen reference basis $K$.
\subsection{Quantum Simulation of Dephasing Dynamics}
We theoretically simulate a single-qubit dephasing noise channel based on a NMR system.
The Hamiltonian of the NMR system comprises two components: the internal Hamiltonian of the system and the control Hamiltonian.
By modulating the control Hamiltonian, we apply a time-dependent Hamiltonian to the system
\begin{eqnarray}
H(t)= \frac{\omega_{k}}{2}\sigma_{z}+\beta_z(t)\sigma_{z},  \label{eq2}
\end{eqnarray}
where $H_{s}=\omega_{k}\sigma_{z}/2$  represents the control Hamiltonian with $\omega_{k}$ denoting the Zeeman energy, and $H_{0}(t)=\beta_z(t)\sigma_{z}$ is the injected transverse relaxation noise Hamiltonian.
Here, $\beta(t)=\alpha\omega_{0}\sum_{j=1}^{J}jF(j)\cos\left(\omega_{j}t+\psi_{j}\right)$ models the random phase modulation noise in the time domain \cite{Soare2014},
with $\alpha$ indicating the noise strength, $\omega_{0}$ the noise based frequency, $\omega_{J}$ the noise cutoff frequency, $\psi_j$ a series of random phases, and $F(j)=j^{\frac{p}{2}-1}$ the noise-type modulation function.
In the Schr\"odinger picture, the evolution operator corresponding to Eq.~(\ref{eq2}) is
\begin{eqnarray}
U(t)=\exp(-i\frac{\omega_{k}t}{2}\sigma_{z})\exp\left[-i\int_{0}^{t}d\tau\beta(\tau)\sigma_{z}\right].
\end{eqnarray}
To perform the quantum simulation, we prepare an ensemble of $N$ identical initial states $\rho(0)$. The ensemble's time evolution under $U(t)$ is given by
\begin{eqnarray}\label{eq:rho-t}
\text{\ensuremath{\rho(t)}}	&=& \frac{1}{N}\sum U(t)\rho(0)U^{\dagger}(t) \\
	&=&\left(\begin{array}{cc}
\rho_{00}(0) & \rho_{01}(0)e^{-i\omega_{k}t}e^{-2\Gamma(t)}\\
\rho_{10}(0)e^{i\omega_{k}t}e^{-2\Gamma(t)} & \rho_{11}(0)   \nonumber
\end{array}\right),
\end{eqnarray}
where $\Gamma(t)$ is the decoherence function, satisfying
\begin{eqnarray}
\Gamma(t)&=&\int_{0}^{t}d\tau_{1}\int_{0}^{t}d\tau_{2}\langle\beta_{z}(\tau_{1})\beta_{z}(\tau_{2})\rangle \nonumber\\
&=&2\alpha^{2}\omega_0^{2}\sum_{j=1}^{J}[jF(j)]^{2}\frac{\sin^{2}\frac{\omega_j t}{2}}{\omega_j^{2}}.  \label{eq:decoherence-function}
\end{eqnarray}
According to the above Eq.~(\ref{eq:rho-t}), it can be seen that the off-diagonal elements of the density matrix decay exponentially with time, while the diagonal elements remain unchanged. The additional phase factor $e^{\pm i\omega_k t}$ in the off-diagonal elements stems from the control Hamiltonian, capturing the coherent oscillatory behavior inherent to the system's quantum dynamics. It is important to emphasize that the exponential decay term governs the essential characteristic of decoherence---the irreversible loss of quantum coherence. In contrast, the phase factor only influences the oscillation frequency of the coherent dynamics and does not alter the coherence decay rate. Thus, our quantum simulation has successfully and effectively reproduced the dynamical behavior of the theoretical model for the dephasing noise channel.

\subsection{Basis-Dependent Coherence Dynamics and Non-Markovianity}
To systematically compare the distinct dynamical behaviors across different reference bases, we summarize the key results in Table~\ref{tab:coherence-summary}. This comparative analysis reveals a crucial dichotomy: while the non-Markovianity measure $\mathcal{N}(\Phi_t)$ is universally determined by the decoherence function $\Gamma(t)$ regardless of the basis choice, the coherence evolution $C^l_K(\rho(t))$ exhibits pronounced basis specificity.
Under the $\sigma_z$ basis, revival occurs exclusively in non-Markovian environments ($\dot{\Gamma}(t) < 0$).
Remarkably, we find that coherence revival can occur even in Markovian regimes when the coherence oscillation ($\omega_k$) competes with and temporarily overcomes decoherence in the $\sigma_x$ or $\sigma_y$ basis. This provides a novel control mechanism independent of environmental memory.

Regarding the detailed computational procedures for the quantum coherence (Eq.~(\ref{eq:coh-z-7-1})-(\ref{eq:coh-y-7-3})) and non-Markovianity measures (Eq.~(\ref{eq:non-markov})) under different reference bases, as summarized in Table \ref{tab:coherence-summary} above, a comprehensive derivation and explanation are provided in Appendix~\ref{sec:Appendix-B}.

\begin{table*}[ht]
\centering
\caption{Summary of coherence dynamics under different reference bases}
\label{tab:coherence-summary}
\renewcommand{\arraystretch}{1.8}  
\begin{tabular}{p{1.8cm} p{3.8cm} p{4.0cm} p{5.5cm}}  
\textbf{Basis} & \textbf{Initial State} $|\Psi(0)\rangle$ & \textbf{Coherence} $C^l_K(\rho(t))$ & \textbf{Non-Markovianity} $\mathcal{N}(\Phi_t)$ \\
\hline
$\sigma_z$ & $\frac{1}{\sqrt{2}}(|0\rangle + e^{i\phi}|1\rangle)$ &
\begin{minipage}{4.0cm}
\centering
\begin{equation*}  
\label{eq:coh-z-7-1}  
e^{-2\Gamma(t)}
\tag{7-1}  
\end{equation*}
\end{minipage}
& \multirow{3}{*}{
\begin{minipage}{5.5cm}
\centering
\begin{equation}
\label{eq:non-markov}
-2 \int_{\dot{\Gamma}<0} \dot{\Gamma}(t) e^{-2\Gamma(t)} dt
\tag{8}  
\end{equation}
\end{minipage}
} \\
$\sigma_x$ & $\frac{1}{\sqrt{2}}(|+\rangle + i|-\rangle)$ &
\begin{minipage}{4.0cm}
\centering
\begin{equation*}
\label{eq:coh-x-7-2}
|\cos(\omega_k t)| e^{-2\Gamma(t)}
\tag{7-2}  
\end{equation*}
\end{minipage}
& \\
$\sigma_y$ & $\frac{1}{\sqrt{2}}(|i+\rangle + i|i-\rangle)$ &
\begin{minipage}{4.0cm}
\centering
\begin{equation*}
\label{eq:coh-y-7-3}
|\cos(\omega_k t)| e^{-2\Gamma(t)}
\tag{7-3}  
\end{equation*}
\end{minipage}
& \\
\hline
\end{tabular}
\end{table*}

\section{RESULTS AND DISCUSSION}\label{sec:Results}
\subsection{Non-Markovianity and Decoherence Function Dynamics}
As evident from Eq.~(\ref{eq:non-markov}) in the Table \ref{tab:coherence-summary}, the non-Markovianity measure is directly determined by the decoherence function $\Gamma(t)$. Our previous work \cite{zhang2025,Long2022,Chen2022} demonstrated that Markovian and non-Markovian environments can be selectively engineered by tuning the noise based frequency $\omega_0$ within specific time intervals. Notably, Ref.~\cite{zhang2025} established that if a noise based frequency $\omega_0$ creates non-Markovian behavior over time $t$, then approximately $\omega_0/2$ is required to maintain non-Markovianity over $2t$. The present work rigorously derives the critical noise based frequency $\omega_0^{c}$ for arbitrary time intervals $t\in[0,t_{\text{max}}]$, providing precise parameter guidelines for non-Markovian environment construction across different timescales.
\begin{figure*}[ht]
\includegraphics[scale=0.5]{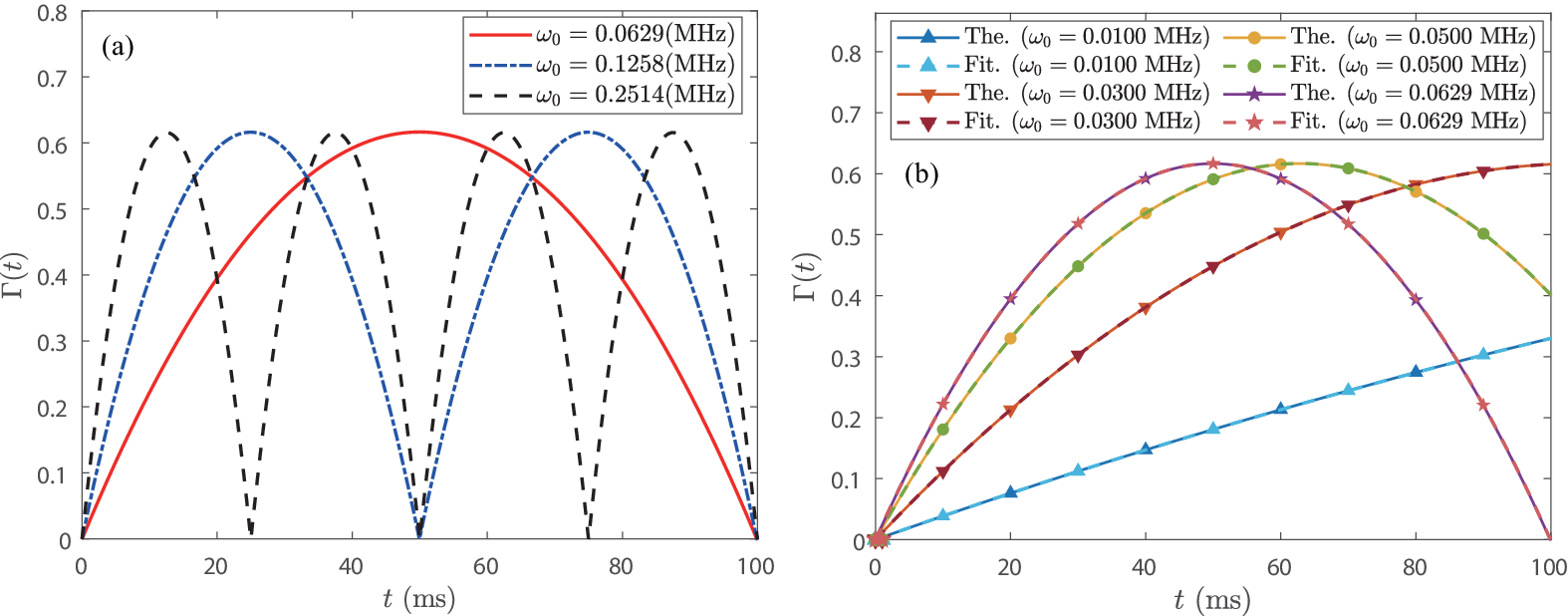}
\centering
\caption{Decoherence function $\Gamma(t)$ dynamics. (a) Time evolution for $\omega_0\in[0.06285,0.1258,0.2514]$~MHz. (b) Comparison between $\Gamma(t)$ and fitted $\Gamma'(t)$ for $\omega_0\in[0.01,0.03,0.05,0.0629]$~MHz. Parameters: $\alpha=0.5$, $\omega_J=50$~MHz.}
\label{fig1}
\end{figure*}
\begin{figure*}
\centering
\includegraphics[width=\linewidth]{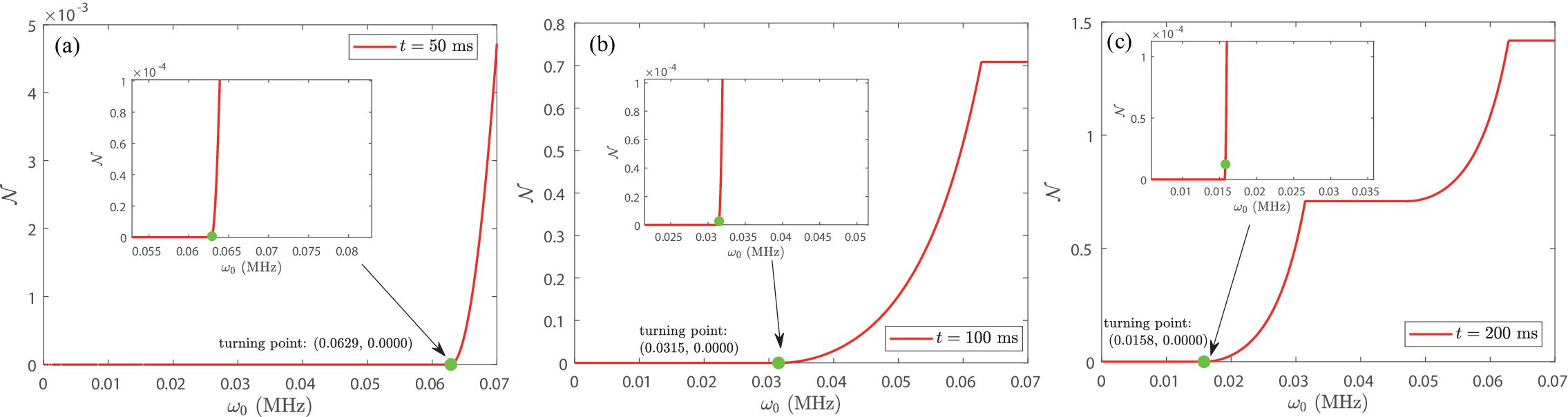}
\caption{(a)-(c) Non-Markovianity measure $\mathcal{N}$ versus noise based frequency $\omega_0$ for time intervals $t\in[0,50]$~ms, $[0,100]$~ms, and $[0,200]$~ms respectively, with $\omega_J=50$~MHz, $\alpha=0.5$, and $F(j)=1/j$. The dynamical phase transitions occur at $\omega_0^c=0.0629$, $0.0315$, and $0.0158$~MHz for the respective time intervals.}
\label{fig2}
\end{figure*}

Equation~(\ref{eq:decoherence-function}) indicates that the decoherence function \(\Gamma(t)\) oscillates periodically with a period determined by \(\omega_0\). Fig.~\ref{fig1}(a) demonstrates this for \(t \in [0,100]\) ms (\(\alpha=0.5\), \(\omega_J=50\) MHz, white noise $F(j)=1/j$): setting \(\omega_0 = 0.06285\) MHz, \(0.1258\) MHz, and \(0.2514\) MHz produces one, two, and three oscillation periods, respectively, generalizing to \(n\) periods for \(\omega_0 = n \cdot 0.06285\) MHz. When the interval is doubled to \(t \in [0,200]\) ms, the required frequency for one period scales to \(0.06285/2\) MHz. This leads to the universal relation \(\omega_0 = n \cdot (6.285 / t_{\text{max}})\) MHz for \(n\) periods in any interval \([0, t_{\text{max}}]\) (see Appendix Fig.~\ref{fig1-SI}).

Since the degree of non-Markovianity in the system is governed by the decoherence function $\Gamma(t)$, we performed a polynomial fit to $\Gamma(t)$ over a single oscillation period $t \in [0, 100]\ \text{ms}$ to support the controlled design of a non-Markovian environment. The fitting procedure was carried out under white noise conditions ($F(j) = 1/j$) with a cutoff frequency $\omega_J = 50\ \text{MHz}$, considering a noise based frequency range of $0 \leq \omega_0 \leq 0.06285\ \text{MHz}$ and a noise strength range of $0 \leq \alpha \leq 1$. The resulting fitted form of the decoherence function is expressed as:
\begin{equation}
\label{eq:decoherence-fit}
\Gamma'(t) = 1.57\alpha^2\omega_0 t - 0.2498\alpha^2\omega_0^2 t^2,
\tag{9}
\end{equation}
As shown in Fig.~\ref{fig1}(b), we compare the theoretical decoherence function (Eq.~(\ref{eq:decoherence-function})) with the fitted function (Eq.~(\ref{eq:decoherence-fit})) for several noise based frequencies $\omega_0 = [0.01, 0.03, 0.05, 0.06285]$ MHz, with fixed $\alpha = 0.5$ and $\omega_J = 50$ MHz. The agreement between the two is excellent. This confirms that the noise based frequency governs the periodicity of the decoherence function, underscoring its critical role in engineering non-Markovian environments.
It is important to note that although the fitted decoherence function in Eq.~(\ref{eq:decoherence-fit}) was obtained with $\omega_J = 50$ MHz, we further demonstrate that variations in the cutoff frequency have negligible impact on the decoherence function. As shown in Fig.~\ref{fig2-SI} of the Appendix, the theoretical and fitted decoherence functions show excellent agreement across a wide range of cutoff frequencies ($\omega_J = [100,200,500,1000,2000,4000]$ MHz).

\begin{figure*}
\centering
\includegraphics[width=\linewidth]{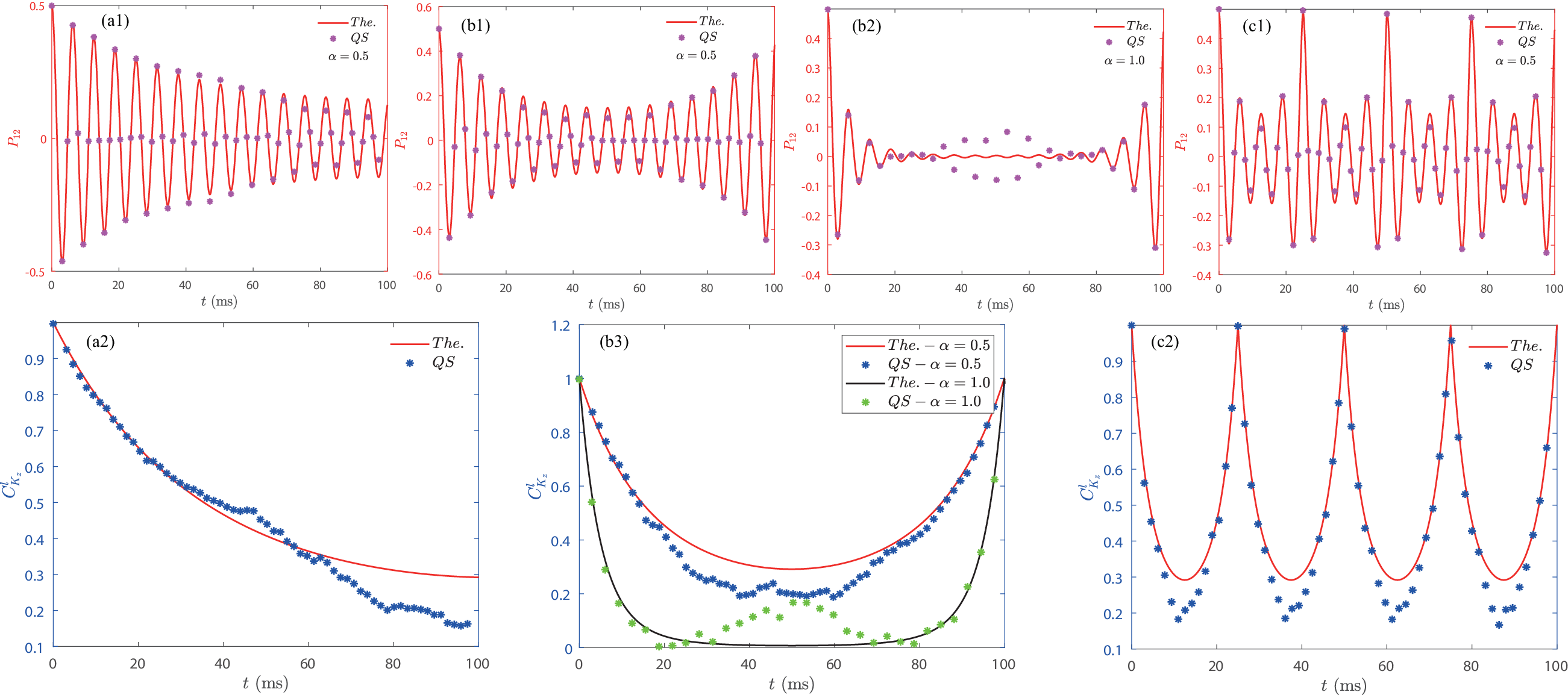}
\caption{Temporal evolution of the off-diagonal density matrix elements and the corresponding quantum coherence. (a1, a2) Dynamics under a noise based frequency of $\omega_0=0.03$~MHz and noise strength $\alpha=0.5$. (b1-b3) Dynamics under $\omega_0=0.0314$~MHz with $\alpha=0.5$ (b1, b3) and $\alpha=1.0$ (b2, b3). (c1, c2) Dynamics under $\omega_0=0.2514$~MHz and $\alpha=0.5$. All quantum simulations were performed with an ensemble size of $N=500$.}
\label{fig3}
\end{figure*}

During $t\in[0,100]$~ms, combining the fitted decoherence function (Eq.~(\ref{eq:decoherence-fit})) with the non-Markovianity measure (Eq.~(\ref{eq:non-markov})), the condition for non-Markovian environment construction requires $-2\dot{\Gamma}(t)e^{-2\Gamma(t)}>0$, yielding the critical phase transition point
\begin{equation}\label{eq:ctirical-condition-N}
\omega_0^{c} = \frac{1.57}{0.4996 \cdot t_s}~\text{MHz},
\tag{10}
\end{equation}
where $t_s$ denotes the earliest time at which $\mathcal{N}>0$. This implies that constructing non-Markovian environments within $t\in[0,t_{\max}]$ requires setting $\omega_0$ such that $t_s \leq t_{\max}$.
When $t_s = t_{\max}$, the corresponding value $\omega_0=\frac{1.57}{0.4996 \cdot t_{\text{max}}}$ is the minimum critical noise based frequency.
For instance, ensuring $t_s\leq100$~ms for $t\in[0,100]$~ms demands $\omega_0>\omega_0^c=0.0314$~MHz, as verified in Fig.~\ref{fig2}(b).
Although the fitted decoherence function (Eq.~(\ref{eq:decoherence-fit})) was obtained for $t\in[0,100]$~ms, the periodicity of the decoherence function with respect to the noise based frequency $\omega_0$, combined with our previous finding \cite{zhang2025} that non-Markovian environment construction requires $\omega_0$ for time interval $t$ but approximately $\omega_0/2$ for $2t$, demonstrates that the critical condition in Eq.~(\ref{eq:ctirical-condition-N}) holds universally for arbitrary time intervals.
We confirm this by analyzing $\mathcal{N}(\Phi_{t})$ for $t\in[0,50]$~ms and $t\in[0,200]$~ms using the theoretical decoherence function (Eq.~(\ref{eq:decoherence-function})). Figures~\ref{fig2}(a) and (c) reveal the globally minimum critical frequencies of $\omega_0^c = 0.0629$ MHz and $0.0158$ MHz, respectively, which correspond to the case of $t_s = t_{\text{max}}$. These values match Eq.~(\ref{eq:ctirical-condition-N}) with an error of only $10^{-4}$.
In addition, the fitted decoherence function (Eq.~(\ref{eq:decoherence-fit})) is independent of the noise cutoff frequency, which makes the critical phase transition point (Eq.~(\ref{eq:ctirical-condition-N})) universal.

Thus, Eq.~(\ref{eq:ctirical-condition-N}) provides an exact, analytical criterion for engineering the dynamical nature of the environment.
For any desired maximum evolution time $t_{\text{max}}$, setting $\omega_0 > \omega_0^c$ guarantees the emergence of non-Markovianity within that interval. This closes the long-standing gap of lacking a precise design rule for constructing non-Markovian environments on demand.
What's more, the decoherence function (Eq.~(\ref{eq:decoherence-function})) implemented via bath-engineering techniques \cite{Soare2014} makes the critical phase transition point $\omega_0^{c}$ applicable across quantum platforms including NMR, superconducting circuits, and ion traps \cite{Khaneja2005}. While the noise strength $\alpha$ doesn't affect the critical condition, it inversely modulates $\mathcal{N}$ magnitude through $\Gamma(t)$, as evident from Eqs.~(\ref{eq:decoherence-function}) and (\ref{eq:decoherence-fit}).

\subsection{Z-Basis Revival as a Non-Markovian Resource}
For the $\sigma_z$ reference basis, starting from the maximal coherent state $|\psi\rangle = (|0\rangle + |1\rangle)/\sqrt{2}$, the coherence after the dephasing channel is described by Eq.~(\ref{eq:coh-z-7-1}) in Table \ref{tab:coherence-summary}, consistent with \cite{zhang2025}. Coherence revival occurs if the decoherence function $\Gamma(t)$ decreases after an initial increase (a non-Markovian signature). However, when $\Gamma(t) \geq 3$, the coherence $C_{K_z}^{l}(\rho(t)) \approx 0.0025$ is effectively zero, meaning revival is unobservable for large decoherence. The noise strength $\alpha$ is the primary factor determining the scale of $\Gamma(t)$.

For the specific time interval $t\in[0,100]$~ms, the minimum critical noise based frequency for constructing a non-Markovian environment is $\omega_0^{c}=\frac{1.57}{0.4996\times100}=0.0314$~MHz.
We examine three cases relative to this threshold:
$\omega_0=0.03$~MHz (below critical, Markovian), $\omega_0=0.06285$~MHz (above critical, non-Markovian), and $\omega_0=0.2514$~MHz (far above critical, non-Markovian).
Their coherence dynamics are shown in Fig.~\ref{fig3}. Panels (a1), (b1-b2), and (c1) display the evolution of the off-diagonal density matrix elements, where red solid curves and purple asterisks represent the theoretical and GRAPE \cite{Khaneja2005} quantum simulation results , respectively.
The corresponding coherence evolution is shown in panels (a2), (b3), and (c2), with solid curves and asterisks (blue/green) again denoting theory and quantum simulation data ($N=500$ ensembles, Zeeman energy $\omega_k=1$~MHz, noise cutoff frequency $\omega_J=50$~MHz).
As expected, monotonic decay occurs in the Markovian regime (a2), while coherence revival is observed in the non-Markovian cases (b3, c2). Notably, a weaker noise strength
$\alpha$ enhances the revival magnitude (b3).
Minor discrepancies between theory and quantum simulation, attributable to the finite ensemble size \cite{Chen2022,Zhang2021}, become more pronounced with increasing $\alpha$; nevertheless, both methods yield fully consistent trends and conclusions.

Furthermore, for $t\in[0,100]$~ms, setting $\omega_0=n\cdot0.06285$~MHz ($n\in\mathbb{Z}$) induces periodic full coherence revivals to unity at $t=n\cdot\frac{6.285}{\omega_0}$~ms, where $n$ determines revival multiplicity. As shown in Fig.~\ref{fig3}(c2) for $\omega_0=0.2514$~MHz ($n=4$), coherence complete revivals occur at $t=25,50,75,100$~ms. This generalizes to arbitrary $t\in[0,t_{\text{max}}]$ when
\begin{equation}\label{eq:critical-point-z}
\omega_0=n\cdot\frac{6.285}{t_{\text{max}}}~\rm{MHz},\quad n\in\mathbb{Z}, \tag{11}
\end{equation}
yielding revival periods $T=\frac{6.285}{\omega_0}$, and revivals at times $t=n\cdot\frac{6.285}{\omega_0}$ ($1\leq n\leq\frac{\omega_0 t_{\text{max}}}{6.285}$). Fig.~\ref{fig3-SI} in Appendix verifies this for $t\in[0,50]$~ms and $t\in[0,200]$~ms.

The $\sigma_{z}$ basis coherence control scheme demonstrates two distinct operational regimes.
First, coherence revival within time interval $t\in[0,t_{\text{max}}]$ is achieved by establishing a non-Markovian environment through $\omega_0>\frac{1.57}{0.4996\cdot t_{\text{max}}}$~MHz, where reduced noise strength $\alpha$ enhances revival amplitude. Second, precisely timed periodic revivals emerge when the noise based frequency follows $\omega_0=n\cdot\frac{6.285}{t_{\text{max}}}$~MHz (with $n\in\mathbb{Z}$), producing revival cycles with period $T=\frac{6.285}{\omega_0}$.
This dual-parameter control strategy enables programmable coherence recovery in quantum systems.

\subsection{Basis-Dependent Coherence Control: Markovian Revival and Noise Filtering}
\begin{figure*}[ht]
\includegraphics[scale=0.5]{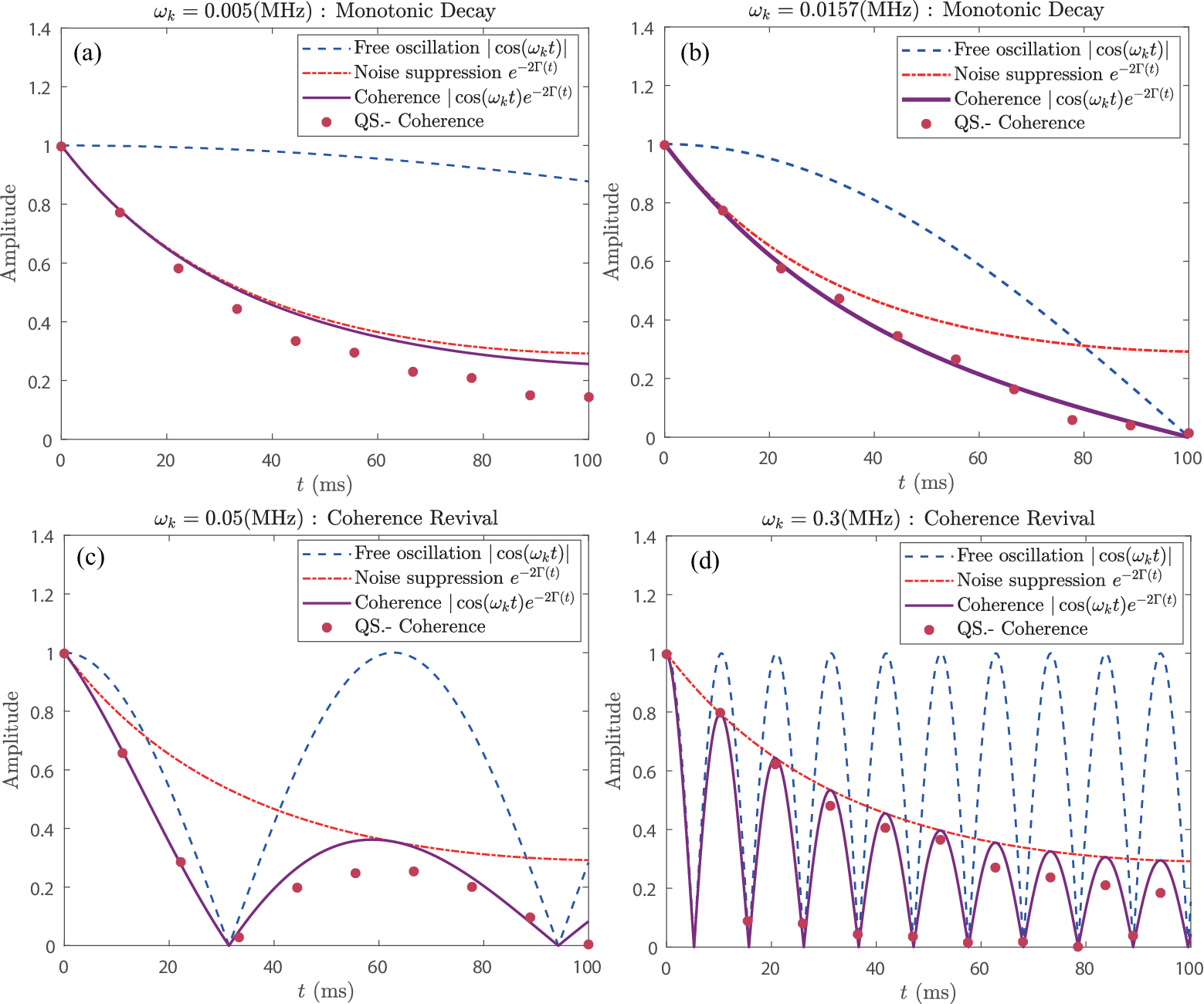}
\centering
\caption{In Markovian environment ($\omega_0=0.03$ MHz): (a-d) Time evolution of free oscillation term $\vert\cos(\omega_{k}t)\vert$, decay term $e^{-2\Gamma(t)}$, and quantum coherence $C_{K_{x}}^{l}(\rho_{X}(t))$ for qubit frequencies (a) $\omega_k=0.005$~MHz, (b) $\omega_k=0.0157$~MHz, (c) $\omega_k=0.05$ MHz, and (d) $\omega_k=0.3$~MHz (noise strength $\alpha=0.5$, noise cutoff frequency $\omega_J=50$~MHz, ensemble size $N=500$). Cases (a) and (b) correspond to $\omega_k\leq\omega_k^{c}$, while (c) and (d) demonstrate $\omega_k>\omega_k^{c}$.}
\label{fig4}
\end{figure*}
Since quantum coherence and non-Markovianity exhibit identical evolutionary behavior under the $\sigma_x$ and $\sigma_y$ reference bases when a maximal coherent state passes through the simulated dephasing channel, we restrict our analysis to the $\sigma_x$ basis as a representative case.
For the $\sigma_x$ basis, the initial maximal coherent state is prepared as $\left|\Psi_x\right\rangle = \frac{1}{\sqrt{2}}(\left|+\right\rangle + i\left|-\right\rangle)$. After evolution through the noise channel, the quantum coherence is described by Eq.~(\ref{eq:coh-x-7-2}), and the non-Markovianity measure is given by Eq.~(\ref{eq:non-markov}) in Table~\ref{tab:coherence-summary}.
Since the non-Markovianity measures for maximal coherent states in the $\sigma_x$/$\sigma_y$ and $\sigma_z$ bases share the same mathematical form after the noise channel, non-Markovian environments can be equally constructed in the $\sigma_x$/$\sigma_y$ bases by tuning the noise based frequency to satisfy $\omega_0 > \omega_0^c = 1.57/(0.4996 \cdot t_{\text{max}})$.

\subsubsection{Markovian environment}
Under the $\sigma_x$ reference basis, the quantum coherence of the maximal coherent state after passing through the noise channel is given by $C_{K_{x}}^{l}(\rho_{X}(t))=\vert\cos(\omega_{k}t)e^{-2\Gamma(t)}\vert$. This expression reveals that besides the decoherence function $\Gamma(t)$, the Zeeman energy $\omega_k$ also significantly influences the quantum coherence behavior, providing a novel approach for coherence control in the $\sigma_x$ basis.

Since the system's non-Markovianity primarily depends on the decoherence function (Eq.~(\ref{eq:non-markov})), coherence revival can still be achieved in Markovian environments by satisfying the extremum condition $\omega_k\tan(\omega_{k}t)=-2\dot{\Gamma}(t)$ (where $\dot{\Gamma}(t)=d\Gamma(t)/dt$) through proper adjustment of the Zeeman energy $\omega_k$. Consequently, we establish the critical Zeeman energy for coherence revival within time $t\in[0,t_{\rm{max}}]$ under the $\sigma_x$ basis as
\begin{equation}
\omega_k^{c}=\frac{\pi}{2\cdot t_{\rm{max}}}~\rm{MHz}. \tag{12}
\end{equation}
Coherence revival occurs when $\omega_k>\omega_k^{c}$, while monotonic decay prevails when $\omega_k\leq\omega_k^{c}$.
\begin{figure*}[ht]
\centering
\includegraphics[scale=0.45]{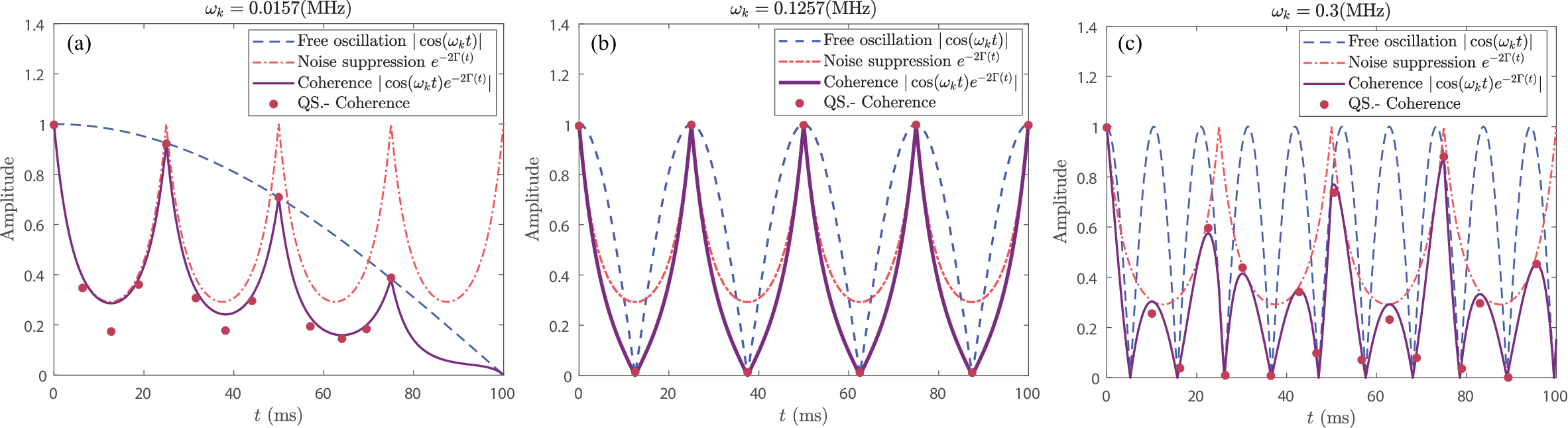}
\caption{Non-Markovian environment ($\omega_0=0.2514$~MHz $>\omega_0^{c}$): Temporal evolution of coherence oscillation term $\vert\cos(\omega_{k}t)\vert$, decay term $e^{-2\Gamma(t)}$, and quantum coherence $C_{K_{x}}^{l}(\rho_{X}(t))$ for qubit frequencies. (a) $\omega_k=0.0157$~MHz, (b) $\omega_k=0.1257=\frac{\pi\omega_0}{6.285}$~MHz, and (c) $\omega_k=0.3$~MHz. The parameters, $\alpha=0.5$, $\omega_J=50$~MHz, $N=500$.}
\label{fig5}
\end{figure*}

Figure \ref{fig4} presents the time evolution over $t \in [0,100]$ ms in a Markovian environment (noise based frequency $\omega_0 = 0.03$ MHz, cutoff frequency $\omega_J = 50$ MHz) for four different Zeeman energies $\omega_k$. The plots show the coherence oscillation term $|\cos(\omega_k t)|$ (blue dashed), the decay term $e^{-2\Gamma(t)}$ (red dash-dotted), and the coherence $C_{K_x}^{l}(\rho_X(t))$ (theoretical results: purple solid line; GRAPE simulations: maroon dots).
When $\omega_k \leq \omega_k^c = \pi / (2 \cdot 100)$ MHz (Figs. \ref{fig4}(a) and (b)), both the coherence oscillation and decay terms decrease monotonically, leading to a monotonic decline in coherence. In contrast, when $\omega_k > \omega_k^c$ (Figs. \ref{fig4}(c) and (d)), the coherence oscillation term increases during certain intervals, surpassing the decay term and resulting in coherence revival.
A comparison across different $\omega_k$ values reveals that in Markovian environments---where the decay term remains fixed---higher Zeeman energies enhance both the magnitude and alter the timing of coherence revivals, owing to the extended oscillation periods of the coherence evolution term.
\begin{figure*}
\includegraphics[scale=0.34]{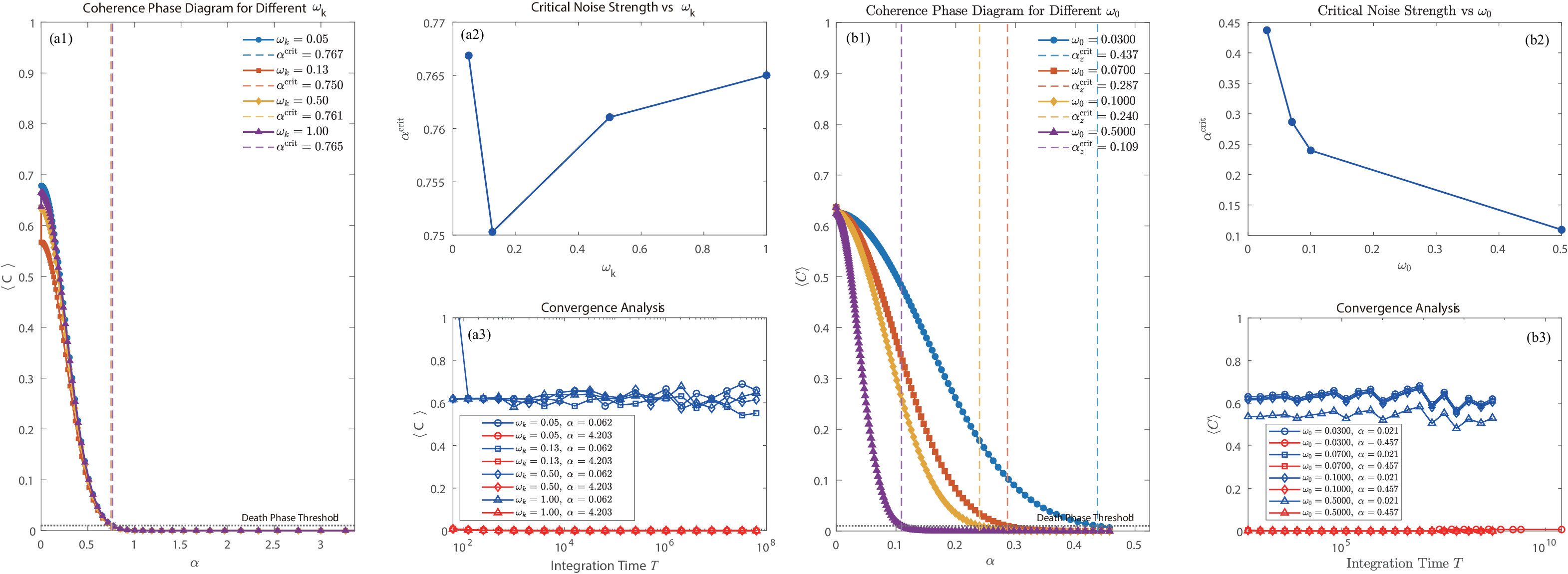}
\centering
\caption{
Long-term coherence averaging under parameter variations. (a1) $\langle C\rangle$ versus $\alpha$ for $\omega_k\in[0.05,0.13,0.5,1.0]$ MHz at fixed $\omega_0=0.03$ MHz. (b1) $\langle C\rangle$ versus $\alpha$ for $\omega_0\in[0.03,0.07,0.1,0.5]$ MHz at fixed $\omega_k=0.1258$ MHz. (a2,b2) Coherence death thresholds $\alpha^{crit}$ versus $\omega_k$ and $\omega_0$ respectively. (a3, b3) Convergence behavior of the time-averaging integral.
 \label{fig6}}
\end{figure*}

Therefore, when selecting the $\sigma_x/\sigma_{y}$ basis as reference and operating within time $t\in[0,t_{\rm{max}}]$ in Markovian environment, adjusting the Zeeman energy to satisfy $\omega_k>\omega_{k}^{c}$ enables quantum coherence revival, where increasing $\omega_k$ enhances both the revival magnitude and timing.

\subsubsection{non-Markovian environment}
Within the time interval $t\in[0,100]$~ms, we set the noise based frequency to $\omega_0=0.2514$~MHz, which exceeds the critical condition $\omega_0^{c}$, establishing a non-Markovian environment.
In this regime, both the decoherence function $\Gamma(t)$ and the decay term $e^{-2\Gamma(t)}$ complete four oscillation cycles.
We analyze three Zeeman energies $\omega_k =[0.0157, 0.1257, 0.3]$ MHz by comparing the coherence oscillation term $|\cos(\omega_k t)|$, the decay term $e^{-2\Gamma(t)}$, and the coherence $C_{K_x}^{l}(\rho_X(t))$.

For $\omega_k = \omega_k^{c}=0.0157$ MHz (Fig.~\ref{fig5}(a)), the coherence decays to zero by $t = 100$ ms but exhibits transient revival during its evolution.
In Fig.~\ref{fig5}(b), when $\omega_k = 0.1257$ MHz exceeds the critical value $\omega_{k}^{c}$ and matches the oscillation period of the decay function, the coherence shows periodic full revival to unity at $t = n \cdot 6.285 / \omega_0$ for $n = 1, 2, 3, 4$, with revival period $T = 6.285 / \omega_0$. In contrast, Fig.~\ref{fig5}(c) illustrates that a mismatch between the coherence oscillation and decay periods results in only partial coherence recovery without full revival.

This phenomenon generalizes to arbitrary time intervals \( t \in [0, t_{\text{max}}] \): when $\omega_0$ is set to \( \omega_0 = n \cdot \frac{6.285}{t_{\text{max}}} \), the decoherence function exhibits \( n \) oscillation cycles. By choosing the Zeeman energy as
\begin{equation}
\omega_k^{r} = \frac{\pi \omega_0}{6.285}~\text{MHz},
\tag{13}
\end{equation}
periodic full coherence revival is achieved with period \( T = \frac{6.285}{\omega_0} \). The revivals occur at times \( t = n \cdot \frac{6.285}{\omega_0} \), where \( 1 \leq n \leq \frac{\omega_0 t_{\text{max}}}{6.285} \) and \( n \in \mathbb{Z} \). These results are further verified in Fig.~\ref{fig4-SI} in the Appendix for \( t \in [0, 50] \) ms and \( t \in [0, 200] \) ms.

Our analysis establishes that universal control over quantum coherence revival--including periodic recovery--in the $\sigma_{x}/\sigma_{y}$ basis is achievable by engineering the Zeeman energy to satisfy $\omega_k > \omega_{k}^{c}$, regardless of whether the environment is Markovian or non-Markovian.
The coherence revival peaks correspond to optimal time windows for quantum operations when environmental decoherence is minimized. For precisely controlled periodic full revival within arbitrary time intervals $t\in[0,t_{\rm{max}}]$, we establish a frequency matching protocol: setting the noise based frequency at $\omega_0=n\cdot\frac{6.285}{t_{\rm{max}}}$ ($n\in\mathbb{Z}$) and synchronizing the Zeeman energy via $\omega_k=\omega_k^{r}=\frac{\pi\omega_0}{6.285}$ generates complete coherence revivals with well-defined periodicity $T=\frac{6.285}{\omega_0}$.

\subsubsection{Long-term Coherence Dynamics Analysis}

We numerically investigate the long-term coherence behavior under the $\sigma_{x}/\sigma_{y}$ reference basis through the time-averaged coherence measure
\begin{equation}
\langle C\rangle = \lim_{T\to\infty}\frac{1}{T}\int_{0}^{T}\vert\cos(\omega_k t)\vert e^{-2\Gamma(t)}dt,
\tag{14}
\end{equation}
which characterizes the system-environment equilibrium dynamics. Remarkably,
Eq.~(\ref{eq:ctirical-condition-N}) reveals that even infinitesimal $\omega_0$ values suffice to establish non-Markovianity in the long-time limit, indicating the system inherently evolves into a non-Markovian regime at extended timescales.
Figure \ref{fig6} systematically examines the parametric dependence of $\langle C\rangle$ on: (i) noise based frequency $\omega_0$, (ii) noise strength $\alpha$, and (iii) Zeeman energy $\omega_k$. Key observations include:
For fixed parameters $\omega_0$, $\omega_k$, and $\omega_J = 50\ \text{MHz}$ (Figs.~\ref{fig6}(a1,b1)), $\langle C \rangle$ remains stable when $\alpha < \alpha^{\text{crit}}$, with enhanced stability at lower $\alpha$ values. Here, we define the coherence death phase by $\langle C \rangle < 0.01$;
Zeeman energy variations (Fig.~\ref{fig6}(a1,a2)) negligibly affect the coherence death threshold $\alpha^{crit}$;
Reduced $\omega_0$ significantly enhances $\alpha^{crit}$ (Figs.~\ref{fig6}(b1,b2)), revealing a universal quantum information preservation strategy in non-Markovian environments;
Convergence analysis (Figs.~\ref{fig6}(a3,b3)) validates the robustness of our long-time averaging protocol.
These results demonstrate that suppressing the noise based frequency ($\omega_0$) substantially enhances the system's noise resilience --- a crucial design principle for robust quantum memory implementations. The identified parametric dependencies provide concrete guidelines for optimizing coherence preservation in practical quantum devices.

\section{CONCLUSION}\label{sec:Conclusion}
We present a comprehensive study of quantum coherence dynamics in engineered dephasing channels, implemented using a quantum simulation algorithm. Our investigation begins by establishing a universal criterion for constructing non-Markovian dynamics through control of the noise based frequency $\omega_0$.
Specifically, for any observation window $t \in [0, t_{\text{max}}]$, we derive a minimum critical noise based frequency $\omega_0^c = 1.57/(0.4996\cdot t_{\text{max}})$ from the white noise condition. Exceeding this value marks the transition to non-Markovian dynamics.

Within this environmental control framework, we observe striking basis-dependent coherence evolution.
In the $\sigma_{z}$ basis, the expected dichotomy emerges where Markovian environments lead to monotonic coherence decay while non-Markovian conditions permit revival phenomena.
Furthermore, by setting the noise based frequency to satisfy $\omega_0 = n \cdot \frac{6.285}{t_{\text{max}}} \;,\quad n \in \mathbb{Z}$, we achieve periodic full coherence revival with period \( T = \frac{6.285}{\omega_0} \), where coherence completely recovers to unity at discrete times $t = n \cdot \frac{6.285}{\omega_0}$ for $1 \leq n\leq\frac{\omega_0 t_{\text{max}}}{6.285}$ .

The behavior becomes even more intriguing when examining the $\sigma_{x}/\sigma_{y}$ basis dynamics, where we observe a previously unreported phenomenon: a dynamical competition between the system's coherence oscillation and noise-induced decay enables coherence revival even in nominally Markovian environments, provided the Zeeman energy satisfies $\omega_k > \omega_{k}^{c} = \pi / (2 t_{\text{max}})$.
This surprising result challenges conventional understanding and suggests that the boundary between Markovian and non-Markovian effects depends critically on the observable being measured. The revival mechanism here is fundamentally distinct from that in the $\sigma_{z}$ basis case, relying on the temporal competition between the system's coherence oscillation and the noise-induced decoherence. Extending to non-Markovian regimes in the $\sigma_{x}/\sigma_{y}$ basis, we find that complete periodic revival requires careful matching of time scales through the condition $\omega_0=n\cdot\frac{6.285}{t_{\rm{max}}}$ and $\omega_k =\omega_{k}^{r}= \frac{\pi\omega_0}{6.285}$.

Taken together, these results advance our understanding of quantum coherence control across multiple fronts. The environmental engineering criteria established here provide experimenters with concrete parameters for generating targeted non-Markovian conditions in diverse physical platforms. The observed basis---dependent revival behavior---particularly the unexpected revival in Markovian environments within the $\sigma_{x}/\sigma_{y}$ basis---suggests new pathways for quantum error mitigation that could be more experimentally feasible than previously assumed. Furthermore, the precise temporal control demonstrated over coherence revivals opens avenues for synchronizing quantum operations with naturally occurring coherence peaks, paving the way for more robust quantum information protocols. These insights not only deepen the fundamental understanding of open quantum systems but also deliver practical strategies for enhancing coherence lifetime in applications spanning quantum sensing and computation.

\begin{acknowledgments}
This work was supported by National Natural Science Foundation of China under Grant Nos. 12505013, 62205042;
Chongqing University of Posts and Telecommunications under Grant Nos. A2022-304, A2022-288, A2024-196;
Science and Technology Research Program of Chongqing Municipal Education Commission under Grant No. KJQN202200603;
Program for the Innovative Talents of Postdoctor of Chongqing under Grant No.2209013344731596.
\end{acknowledgments}
\appendix

\section{Dephasing channel model in different bases}\label{sec:Appendix-A}
The Kraus operators corresponding to the dephasing noise channel model are
\begin{eqnarray}
M_{0}&=&\sqrt{1-p}\begin{pmatrix}
1 & 0\\
0 & 1
\end{pmatrix},
M_{1}=\sqrt{p}\begin{pmatrix}
1 & 0\\
0 & 0
\end{pmatrix},
M_{2}=\sqrt{p}\begin{pmatrix}
0 & 0\\
0 & 1
\end{pmatrix}. \nonumber
\end{eqnarray}
A single-qubit quantum state $\overrightarrow{r}=(r_{x},r_{y},r_{z})$ transforms under the dephasing noise channel as
\begin{equation}
\xi(\overrightarrow{r})=\left((1-p)r_{x}, (1-p)r_{y}, r_{z}\right).
\end{equation}
We consider inputting maximally coherent states under different reference bases $\sigma_{x}/\sigma_{y}$ and $\sigma_z$, the quantum coherence after passing through the dephasing noise channel satisfies \cite{Baumgratz2014}
\begin{equation}
C_{k_{x,y,z}}^{l}(\xi(\overrightarrow{r}))=|1-p|.
\end{equation}
Thus, the coherence magnitude of the single-qubit dephasing noise channel is determined by the parameter $p$, which is related to the type of environment in the noise channel. For example, the dephasing process of a qubit system can be described by a microscopic Hamiltonian
\begin{eqnarray}
H\!=\!\frac{1}{2}\omega_a\sigma_z+\sum_{k}\omega_{k}b_{k}^{\dagger}b_{k}+\sum_{k}\sigma_{z}(g_{k}b_{k}^{\dagger}+g_{k}^{*}b_{k}).
\end{eqnarray}
In the energy eigenstate $\sigma_z$ basis, this phase decay process leads to the evolution of the qubit system as
\begin{eqnarray}
\rho(t)
&=&\begin{pmatrix}
\rho_{00} & \rho_{01}e^{-\gamma(t)}\\
\rho_{10}e^{-\gamma(t)} & \rho_{11}   \nonumber
\end{pmatrix},
\end{eqnarray}
where $\gamma(t)$ is decoherence function.

\section{Coherence and non-Markovianity Measures}
\label{sec:Appendix-B}
In the $\sigma_{z}$ reference basis, the eigenbasis is $\{\vert0\rangle,\vert1\rangle\}$, and the input maximal coherent state is
$\vert\psi(0)\rangle=\frac{1}{\sqrt{2}}\left(\vert0\rangle+e^{i\phi}\vert1\rangle\right)$.
After the quantum state passes through the noise channel constructed by quantum simulation, it evolves as
\begin{eqnarray}
\rho(t)	
	&=&\frac{1}{2}\left(\begin{array}{cc}
1 & e^{-i\phi}e^{-i\omega_{k}t}e^{-2\Gamma(t)}\\
e^{i\phi}e^{i\omega_{k}t}e^{-2\Gamma(t)} & 1
\end{array}\right).
\end{eqnarray}

Further using the $l_1$ norm to measure the coherence of the quantum state is \cite{Baumgratz2014}
\begin{eqnarray}
C_{K_{z}}^{l}(\text{\ensuremath{\rho(t)}})=e^{-2\Gamma(t)}.  \label{eq:z-coherence}
\end{eqnarray}
Using the method proposed by Breuer-Laine-Piilo (BLP), the non-Markovianity of the system's dynamic evolution process is \cite{Breuer2009}
\begin{eqnarray}
\mathcal{N}(\Phi_{t})
&=&-2\int_{\sigma(t)>0}\dot{\Gamma}(t)e^{-2\Gamma(t)}dt.   \label{eq:non-markovian-z}
\end{eqnarray}

When selecting the $\sigma_x$ basis as reference, we prepare the initial quantum state as a maximally coherent state $\left|\Psi(0)\right\rangle =\frac{1}{\sqrt{2}}(\left|+\right\rangle +e^{i\phi}\left|-\right\rangle)$. Transforming to the computational basis $\{\vert0\rangle,\vert1\rangle\}$ and evolving through the simulated dephasing channel yields
\begin{equation}
\rho(t) = \frac{1}{2}\begin{pmatrix}
1+\cos\phi & i\sin\phi e^{-i\omega_{k}t}e^{-2\Gamma(t)}\\
-i\sin\phi e^{i\omega_{k}t}e^{-2\Gamma(t)} & 1-\cos\phi   \nonumber
\end{pmatrix}.
\end{equation}
Transforming back to the $\sigma_x$ eigenbasis gives
\begin{eqnarray}\label{fig:matrix}
    &\rho_{X}(t) \nonumber \\
    &=\frac{1}{2}\begin{pmatrix}
      1+\sin\phi\sin(\omega_{k}t)e^{-2\Gamma(t)} & \cos\phi-i\sin\phi\cos(\omega_{k}t)e^{-2\Gamma(t)}\\ \nonumber
      \cos\phi+i\sin\phi\cos(\omega_{k}t)e^{-2\Gamma(t)} & 1-\sin\phi\sin(\omega_{k}t)e^{-2\Gamma(t)}
      \end{pmatrix}, \nonumber
\end{eqnarray}
yielding the quantum coherence measure
\begin{equation}
C_{K_{x}}^{l}(\rho_{X}(t)) = \sqrt{\cos^{2}\phi+\sin^{2}\phi\cos^{2}\omega_{k}t e^{-4\Gamma(t)}}.
\end{equation}
The coherence clearly depends on the phase factor $\phi$. For minimal coherence, we set $\phi=\frac{\pi}{2}+n\pi$ ($n\in\mathbb{Z}$), corresponding to the input state $\left|\Psi\right\rangle =\frac{1}{\sqrt{2}}(\left|+\right\rangle +i\left|-\right\rangle)$. The coherence then simplifies to
\begin{equation}\label{eq:coherence-x}
C_{K_{x}}^{l}(\rho_{X}(t)) = \vert\cos\omega_k t\vert e^{-2\Gamma(t)}.
\end{equation}

With $\sigma_y$ selected as the reference basis and the initial state prepared as the maximal coherent state \( \left|\Psi\right\rangle = \frac{1}{\sqrt{2}}\left( \left|i+\right\rangle + e^{i\phi}\left|i-\right\rangle \right) \), we first expand this state in the computational basis. The state then evolves under the dephasing noise channel into the following form:
\begin{eqnarray}\label{eq5}
\rho(t)	
&=&\frac{1}{2}\begin{pmatrix}
1+\cos\phi & \sin\phi e^{-i\omega_{k}t}e^{-2\Gamma(t)}\\
\sin\phi e^{i\omega_{k}t}e^{-2\Gamma(t)} & 1-\cos\phi   \nonumber
\end{pmatrix}.
\end{eqnarray}
Further transforming it to the \( \sigma_y \) eigenbasis gives
\begin{eqnarray}
&\rho_{Y}(t) \nonumber\\
&=\frac{1}{2}\begin{pmatrix}
1+\sin\phi\sin(\omega_{k}t)e^{-2\Gamma(t)} & \cos\phi-i\sin\phi\cos(\omega_{k}t)e^{-2\Gamma(t)}\\ \nonumber
\cos\phi+i\sin\phi\cos(\omega_{k}t)e^{-2\Gamma(t)} & 1-\sin\phi\sin(\omega_{k}t)e^{-2\Gamma(t)}
\end{pmatrix}. \nonumber
\end{eqnarray}
The quantum coherence of the system is then
\begin{eqnarray}
C_{K_{y}}^{l}(\rho_{Y}(t))=\cos^{2}\phi+\sin^{2}\phi\cos^{2}\omega_{k}t \, e^{-4\Gamma(t)}.
\end{eqnarray}
Consistent with the analysis in the \(\sigma_x\) basis, the minimal quantum coherence is studied by setting \(\phi = \frac{\pi}{2} + n\pi\) for \(n \in \mathbb{Z}\), which simplifies the coherence expression to
\begin{eqnarray}
C_{K_{y}}^{l}(\rho_{Y}(t)) = \vert\cos(\omega_k t)\vert e^{-2\Gamma(t)}.
\end{eqnarray}
This result confirms that the quantum coherence displays identical dynamical behavior under the \(\sigma_x\) and \(\sigma_y\) reference bases.

According to the non-Markovianity measure proposed by Breuer, Laine and Piilo (BLP), the non-Markovian behavior of the system's evolution at time $t$ is given by
$\mathcal{N}=\int_{\sigma(t)>0}\sigma(t)dt$,
where $\sigma(t)=dD(\rho_1(t),\rho_2(t))/dt$,
and $D(\rho_1(t),\rho_2(t)) =\frac{1}{2}\vert\vert\rho_1(t)-\rho_2(t)\vert\vert$ is the trace distance between $\rho_1(t)$ and $\rho_2(t)$, and $|A|=\sqrt{AA^\dagger}$ \cite{Breuer2009}.

In the $\sigma_x$ basis, we choose initial states $\vert\psi_{x1}(0)\rangle=\frac{1}{\sqrt{2}}(\vert+\rangle+i\vert-\rangle)$ and $\vert\psi_{x2}(0)\rangle=\frac{1}{\sqrt{2}}(\vert+\rangle-i\vert-\rangle)$. The initial trace distance is
$D(\rho_{1x}(0),\rho_{2x}(0))=\frac{1}{2}\Vert \rho_{1x}(0)-\rho_{2x}(0)\Vert=\frac{1}{2}\rm{tr}\vert\rho_{1x}(0)-\rho_{2x}(0)\vert=1$.
Under the Hamiltonian $H_X(t)=U_{z\to x}H(t)U_{z\to x}^{\dagger}$,
($U_{z\to x}=\frac{1}{\sqrt{2}}\begin{pmatrix}1 & 1\\1 & -1\end{pmatrix}$, $H(t)=\frac{\omega_{k}}{2}\sigma_{z}+\beta_z(t)\sigma_{z}$), the states evolve to
\begin{eqnarray}
\rho_{1x}(t)
\!=\!\frac{1}{2}\begin{pmatrix}
1+\sin(\omega_{k}t)e^{-2\Gamma(t)} & -i\cos(\omega_{k}t)e^{-2\Gamma(t)}\\ \nonumber
i\cos(\omega_{k}t)e^{-2\Gamma(t)} & 1-\sin(\omega_{k}t)e^{-2\Gamma(t)} \\ \nonumber
\end{pmatrix},\\
\rho_{2x}(t)
\!=\!\frac{1}{2}\begin{pmatrix}
1-\sin(\omega_{k}t)e^{-2\Gamma(t)} & i\cos(\omega_{k}t)e^{-2\Gamma(t)}\\ \nonumber
-i\cos(\omega_{k}t)e^{-2\Gamma(t)} & 1+\sin(\omega_{k}t)e^{-2\Gamma(t)} \\
\end{pmatrix}.
\end{eqnarray}
The trace distance at time $t$ becomes
\begin{eqnarray}
D(\rho_{1x}(t),\rho_{2x}(t))&=&\frac{1}{2}\Vert \rho_{1x}(t)-\rho_{2x}(t)\Vert \\ \nonumber
&=&\frac{1}{2}\rm{tr}\vert\rho_{1x}(t)-\rho_{2x}(t)\vert=e^{-2\Gamma(t)}.
\end{eqnarray}
Thus, $\sigma(t)=-2\dot{\Gamma}(t)e^{-2\Gamma(t)}$, and the non-Markovianity at time $t$ is
\begin{eqnarray}
\mathcal{N}(\Phi_{t})
&=&-2\int_{\sigma(t)>0}\dot{\Gamma}(t)e^{-2\Gamma(t)}dt.      \label{eq:non-markovian}
\end{eqnarray}

In the $\sigma_{y}$ basis, we choose initial states $\vert\psi_{y1}(0)\rangle=\frac{1}{\sqrt{2}}(\vert i+\rangle+i\vert i-\rangle)$ and $\vert\psi_{y2}(0)\rangle=\frac{1}{\sqrt{2}}(\vert i+\rangle-i\vert i-\rangle)$.
The initial trace distance is:
$D(\rho_{1y}(0),\rho_{2y}(0))=\frac{1}{2}\Vert \rho_{1y}(0)-\rho_{2y}(0)\Vert=\frac{1}{2}\rm{tr}\vert\rho_{1y}(0)-\rho_{2y}(0)\vert=1$.
Under the Hamiltonian $H_Y(t)=U_{z\to y}H(t)U_{z\to y}^{\dagger}$,
($U_{z\to y}=\frac{1}{\sqrt{2}}\begin{pmatrix}1 & 1\\i & -i\end{pmatrix}$, $H(t)=\frac{\omega_{k}}{2}\sigma_{z}+\beta_z(t)\sigma_{z}$), the states evolve to
\begin{eqnarray}
\rho_{1y}(t)
=\frac{1}{2}\begin{pmatrix}
1+\sin(\omega_{k}t)e^{-2\Gamma(t)} & -i\cos(\omega_{k}t)e^{-2\Gamma(t)}\\ \nonumber
i\cos(\omega_{k}t)e^{-2\Gamma(t)} & 1-\sin(\omega_{k}t)e^{-2\Gamma(t)}
\end{pmatrix}, \\ \nonumber
\rho_{2y}(t)
=\frac{1}{2}\begin{pmatrix}
1-\sin(\omega_{k}t)e^{-2\Gamma(t)} & i\cos(\omega_{k}t)e^{-2\Gamma(t)}\\ \nonumber
-i\cos(\omega_{k}t)e^{-2\Gamma(t)} & 1+\sin(\omega_{k}t)e^{-2\Gamma(t)}\\
\end{pmatrix}.
\end{eqnarray}
This result is consistent with that under the $\sigma_x$ basis. And the non-Markovianity at time $t$ is
\begin{eqnarray}
\mathcal{N}(\Phi_{t})
&=&-2\int_{\sigma(t)>0}\dot{\Gamma}(t)e^{-2\Gamma(t)}dt.      \label{eq:non-markovian}
\end{eqnarray}

It can be observed that under the $\sigma_x/\sigma_y$ basis, the non-Markovianity of the system exhibits the same variation pattern as under the $\sigma_z$ basis, which is primarily influenced by the decoherence function $\Gamma(t)$.

\section{Universality of the Decoherence Function}
\label{sec:Appendix-C}
\begin{figure*}[ht]
\begin{center}
\includegraphics[scale=0.45]{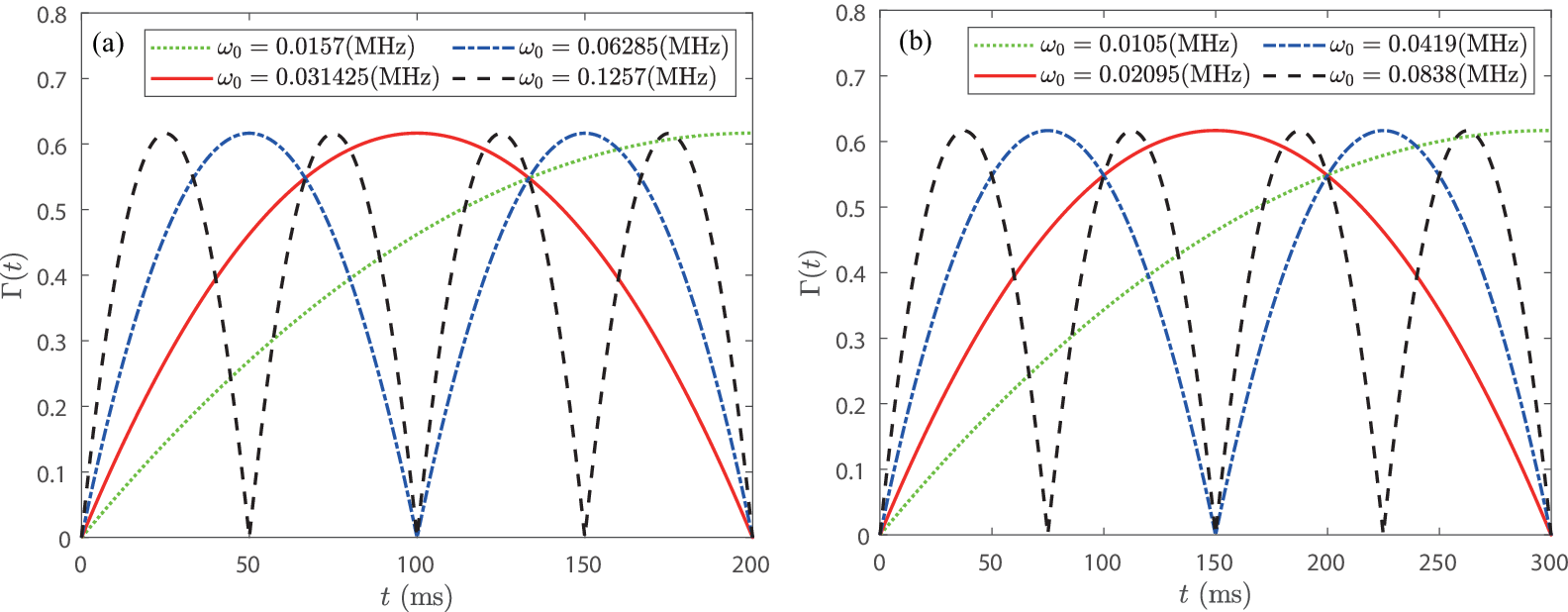}
\caption{Periodic variation of the decoherence function with the noise based frequency \( \omega_0 \) for two different time ranges.}
\label{fig1-SI}
\end{center}
\end{figure*}
\begin{figure*}
\begin{center}
\includegraphics[scale=0.4]{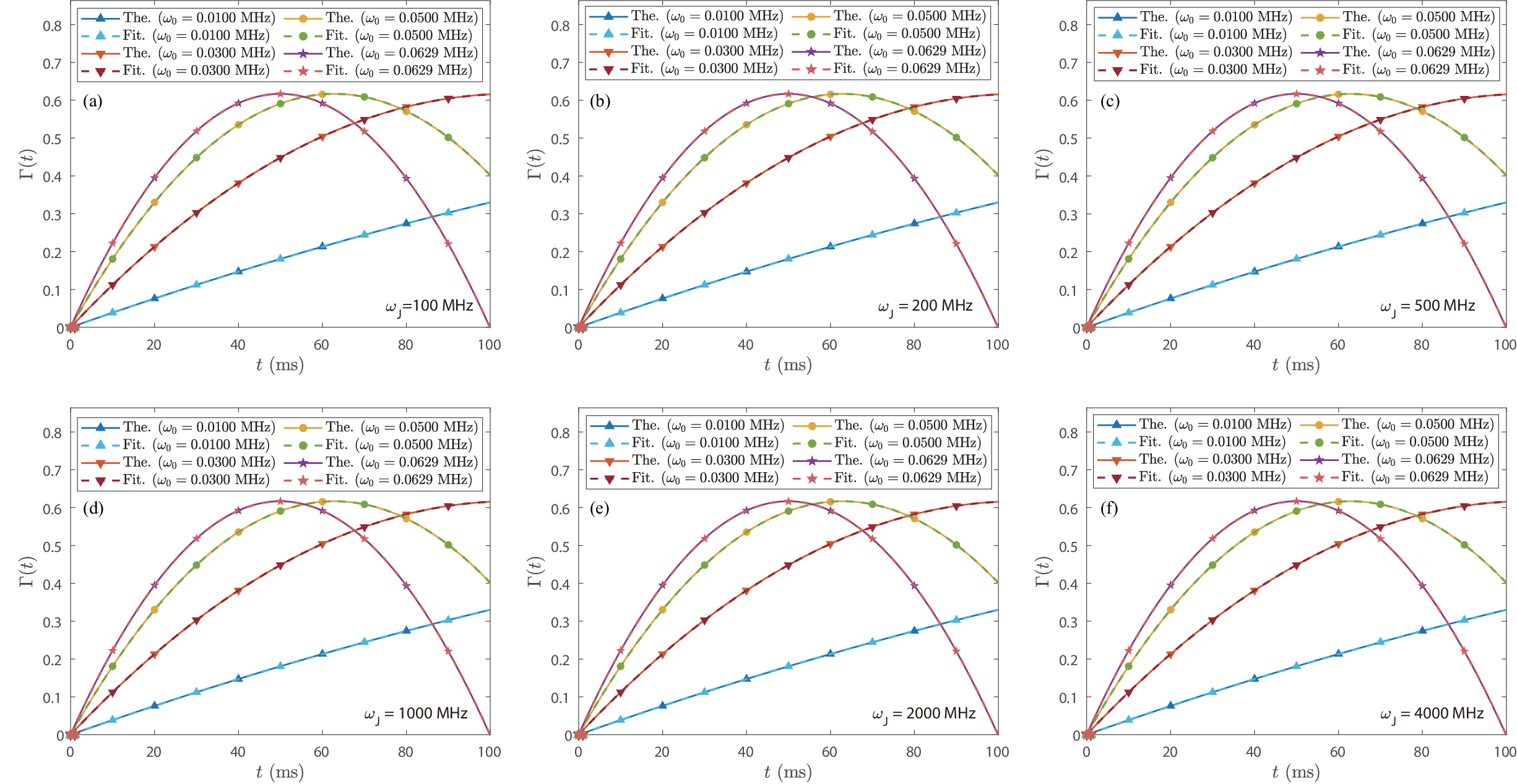}
\caption{Comparison between the fitted decoherence function (Eq.~(\ref{eq:decoherence-fit}) in the main text) and the theoretical decoherence function (Eq.~(\ref{eq:decoherence-function}) in the main text) under different noise cutoff frequencies. Panels (a)-(f) correspond to cutoff frequencies of $\omega_J = 100$, $200$, $500$, $1000$, $2000$, and $4000$ MHz, respectively, with a fixed noise strength of $\alpha = 0.5$.}
\label{fig2-SI}
\end{center}
\end{figure*}
Analysis of the decoherence function in Eq.~(\ref{eq:decoherence-function}) shows that it exhibits periodic oscillations under a fixed noise cutoff frequency, with the oscillation period determined by the noise based frequency over a given time range. In the main text, for \( t \in [0,100]~\text{ms} \) and \( \omega_0 = 0.06285~\text{MHz} \), the decoherence function completes exactly one oscillation period.

As illustrated in Fig.~\ref{fig1-SI}(a), when \( t \in [0,200]~\text{ms} \) and \( \omega_0 = 0.06285/2~\text{MHz} \), one full oscillation period is also completed. For \( \omega_0 = n \times (0.06285/2)~\text{MHz} \) with \( n = 0.5, 1, 3, 4 \), the function completes \( 0.5, 1, 3,\) and \( 4 \) periods, respectively. Similarly, for \( t \in [0,300]~\text{ms} \) and \( \omega_0 = 0.06285/3~\text{MHz} \), one period is completed; more generally, for \( \omega_0 = n \times (0.06285/3)~\text{MHz} \), the function completes \( n \) periods (Fig.~\ref{fig1-SI}(b)).

These observations lead to the general rule: for any interval \( t \in [0, t_{\text{max}}] \), the decoherence function completes one oscillation period when
$\omega_0 = \frac{6.285}{t_{\text{max}}}~\text{MHz}$,
and completes \( n \) periods when
$\omega_0 = n \cdot \frac{6.285}{t_{\text{max}}}~\text{MHz}$.

Furthermore, although the fitted decoherence function (Eq.~(\ref{eq:decoherence-fit})) was obtained under \( \omega_J = 50~\text{MHz} \) for \( t \in [0,100]~\text{ms} \), we find that the cutoff frequency \( \omega_J \) has no appreciable effect on the fitted result. As shown in Fig.~\ref{fig2-SI}, excellent agreement is maintained between the fitted and theoretical decoherence functions (Eqs.~(\ref{eq:decoherence-function}) and (\ref{eq:decoherence-fit}) in the main text, respectively) across a wide range of cutoff frequencies--specifically, \( \omega_J = 100, 200, 500, 1000, 2000, 4000~\text{MHz} \) (panels (a)-(f)).

This consistency confirms that the critical noise based frequency for constructing non-Markovian environments (Eq.~(\ref{eq:ctirical-condition-N}) in the main text) is universal with respect to the noise cutoff frequency.

\section{Revival Conditions Across Bases} \label{sec:Appendix-D}
\subsection{$\sigma_z$ basis}
\begin{figure*}
\begin{center}
\includegraphics[scale=0.45]{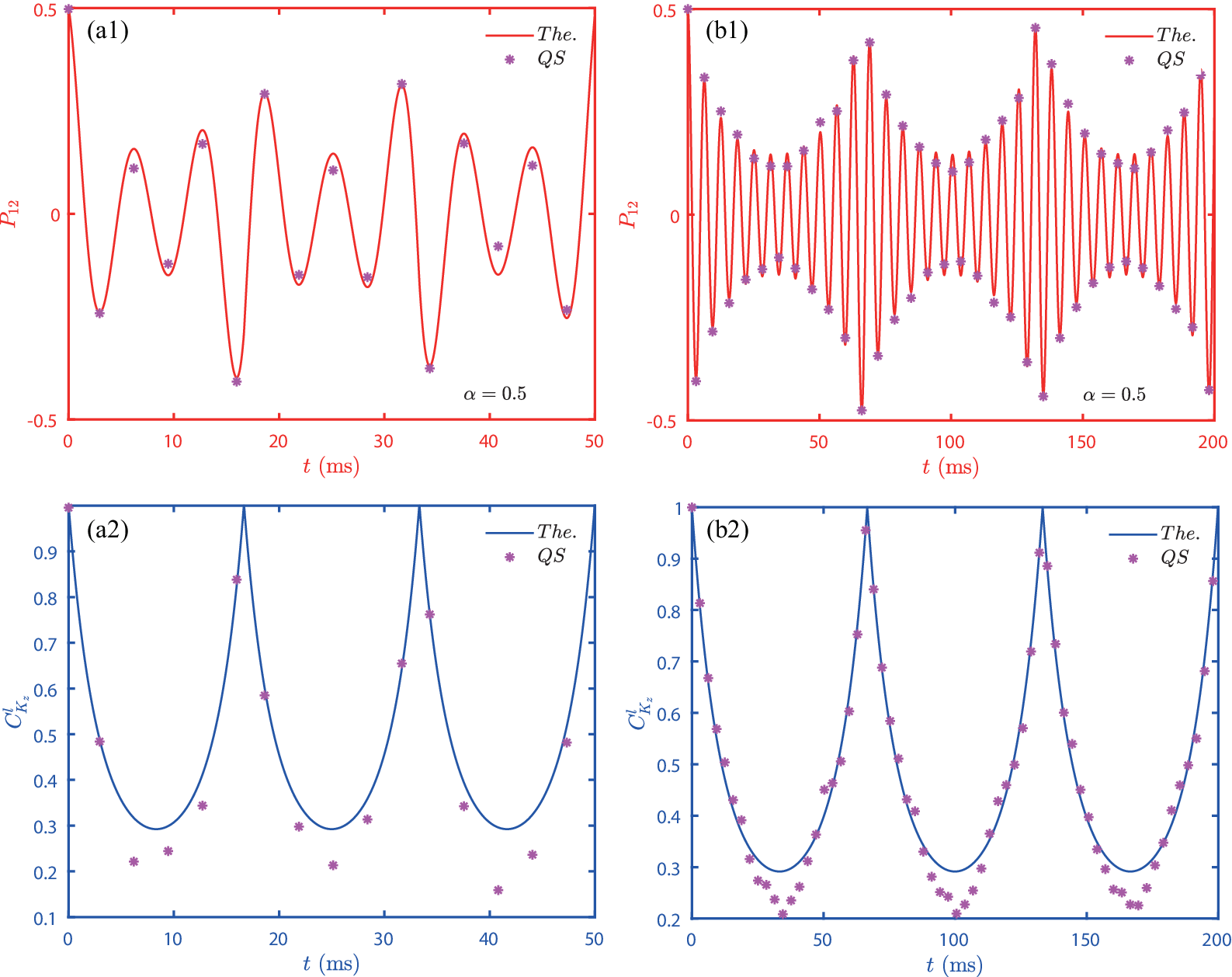}
\caption{Panels (a1) and (a2) show the time evolution of the off-diagonal density matrix element \( P_{12}(t) \) and the quantum coherence \( C_{k_{z}}^{l}(\rho(t)) \), respectively, for \( t \in [0, 50] \) ms. Correspondingly, panels (b1) and (b2) display the same quantities over the extended interval \( t \in [0, 200] \) ms. In all plots, solid curves represent theoretical results, and dots indicate quantum simulations obtained via the GRAPE algorithm ($N=500$ ensembles).}
\label{fig3-SI}
\end{center}
\end{figure*}
\begin{figure*}
\begin{center}
\includegraphics[scale=0.5]{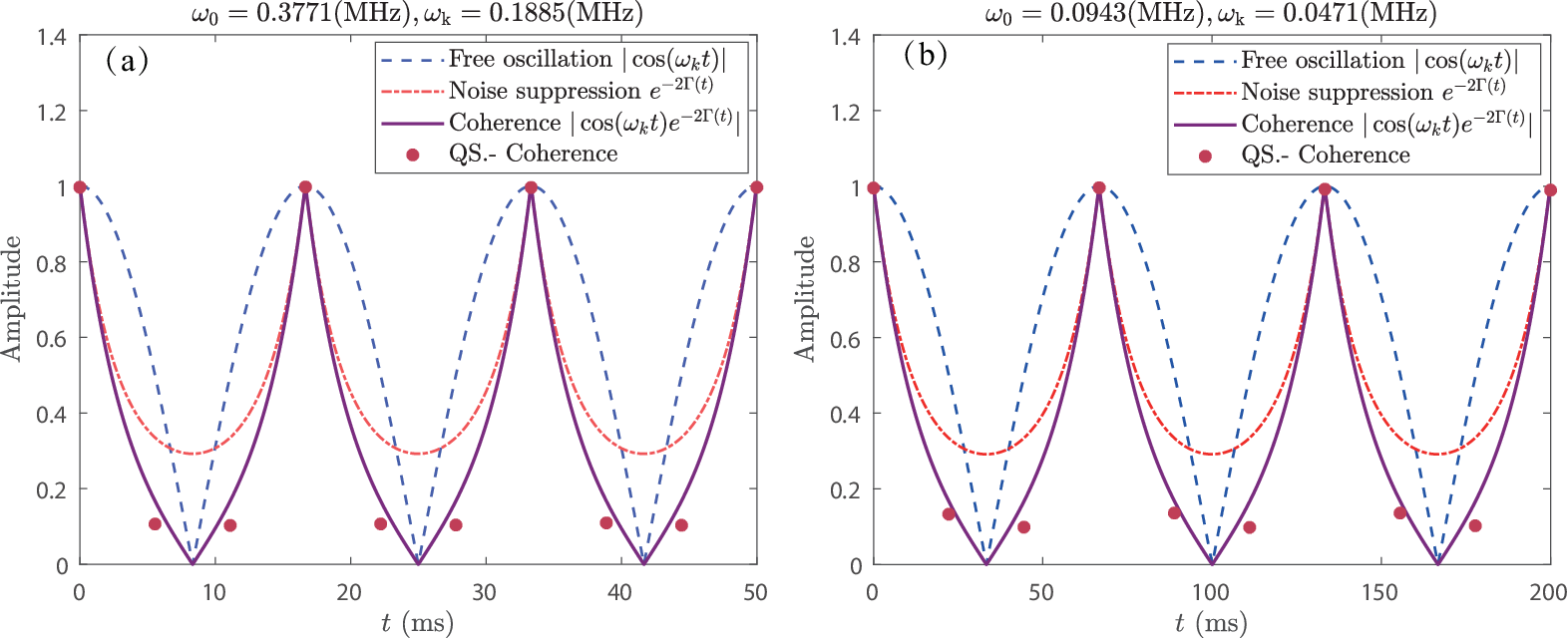}
\caption{Periodic complete revival of quantum coherence under the $\sigma_x$ basis. Panels (a) and (b) correspond to the time intervals \( t \in [0,50] \) ms and \( t \in [0,200] \) ms, respectively. The parameters are \( \omega_0 = 3 \cdot {6.285}/{t_{\text{max}}} \), \( \omega_k = {\pi \omega_0}/{6.285} \), \( \alpha = 0.5 \), \( \omega_J = 50 \) MHz, and the ensemble size \( N = 500 \).}\label{fig4-SI}
\end{center}
\end{figure*}
In the main text, we demonstrated that under the $\sigma_z$ basis, quantum coherence of the system exhibits monotonic decay in a Markovian environment but revives in a non-Markovian environment. Furthermore, by setting the noise based frequency as
$\omega_0=\frac{n\cdot6.285}{t_{\rm{max}}}\,~\rm{MHz},\ n\in\mathbb{Z}$,
periodic complete revival of coherence can be achieved, with the corresponding revival period being $T=\frac{6.285}{\omega_{0}}$ and the revivals occurring at $t=n\cdot\frac{6.285}{\omega_{0}},1\leq n\leq\frac{\omega_0 t_{\rm{max}}}{6.285}$ and $n\in\mathbb{Z}$.

We further verify this conclusion within $t\in[0,50]$~ms and $t\in[0,200]$~ms.
For $t\in[0,50]$~ms, with $\omega_0=3\cdot\frac{6.285}{50}=0.3771$~MHz, $\omega_J=50$~MHz, $\alpha=0.5$, and $\omega_k=1$~MHz, the time evolution of the off-diagonal elements of the density matrix is shown in Fig.~\ref{fig3-SI} (a1), and that of quantum coherence is shown in Fig.~\ref{fig3-SI} (a2), where solid lines represent the results of theoretical calculations, and dots represent the results of quantum simulations ($N=500$ ensembles).
Notably, the coherence exhibits three complete periodic revivals to unity (\(C_{K_{z}}^{l}(\rho(t))=1\)), with revival times precisely matching the theoretical prediction \(t = n \cdot 6.285/\omega_0\) (\(n \in \mathbb{Z}\)). This confirms the critical role of noise based frequency \(\omega_0\) in governing revival periodicity.

When extending the observation window to \( t \in [0,200] \) ms with proportionally scaled \(\omega_0 =3\cdot\frac{6.285}{200}=0.0943\) MHz (maintaining \(\omega_J = 50\) MHz, \(\alpha = 0.5\), and \(\omega_k = 1\) MHz), the off-diagonal elements and quantum coherence evolve as shown in Figs.~\ref{fig3-SI}(b1) and \ref{fig3-SI}(b2), respectively, demonstrating three complete coherence revivals to \(C_{K_{z}}^{l}(\rho(t))=1\) occur at temporally equidistant points, again adhering to the \(6.285/\omega_0\) periodicity. The consistent revival behavior across both time regimes provides validation of Eq.~(\ref{eq:critical-point-z})'s universal validity for arbitrary time conditions, establishing it as a general design rule for non-Markovian coherence control.

\subsection{$\sigma_x/\sigma_y$ basis}
In the main text, we demonstrate that quantum coherence revives in non-Markovian environments under the \(\sigma_x/\sigma_y\) reference bases. By setting the noise based frequency to \(\omega_0 = n \cdot \frac{6.285}{t_{\text{max}}}\) (\(n \in \mathbb{Z}\)) and the Zeeman energy to \(\omega_k = \omega_k^r = \frac{\pi \omega_0}{6.285}\), periodic quantum coherence recovery is achieved. The revival period is \(T = \frac{6.285}{\omega_0}\), with revivals occurring at times \(t = n \cdot \frac{6.285}{\omega_0}\), where \(n \in \mathbb{Z}\) and \(1 \leq n \leq \frac{\omega_0 t_{\text{max}}}{6.285}\).

We further verify this within $t\in[0,50]$~ms and $t\in[0,200]$~ms.
For $t\in[0,50]$~ms, with $\omega_0=3\cdot\frac{6.285}{50}=0.3771$~MHz, $\omega_J=50$~MHz, $\alpha=0.5$, and $\omega_k=0.1885$~MHz, the time evolutions of the coherence oscillation term $\vert\cos(\omega_k t)\vert$, decay term $e^{-2\Gamma(t)}$, and quantum coherence are shown in Fig.~\ref{fig4-SI}(a), exhibiting three complete periodic revivals to unity (\(C_{K_{x}}^{l}(\rho_{X}(t))=1\)).
For $t\in[0,200]$~ms, with $\omega_0=3\cdot\frac{6.285}{200}=0.0943$~MHz, $\omega_J=50$~MHz, $\alpha=0.5$, and $\omega_k=0.0471$~MHz, the coherence evolutions are shown in Fig.~\ref{fig4-SI}(b), also exhibiting three complete revivals.
These results further demonstrate the universality of the condition for periodic full coherence revival under the \(\sigma_x\) and \(\sigma_y\) reference bases---namely, \(\omega_0 = n \cdot \frac{6.285}{t_{\text{max}}}\) (\(n \in \mathbb{Z}\)) and \(\omega_k = \omega_k^r = \frac{\pi \omega_0}{6.285}\)---over arbitrary time intervals.

\section{GRAPE algorithm}\label{sec:Appendix-E}
Experimental simulation of noise channels in NMR systems comprises three key steps: preparing a pseudopure state, evolving it under a random noise Hamiltonian, implementing the evolution operator $U$ via the gradient ascent pulse engineering (GRAPE) algorithm, and extracting information from the final state using quantum state tomography. Detailed procedures are described in our previous studies \cite{Chen2022,Long2022,Wang2018,zhang2025}; here, we briefly introduce the GRAPE algorithm.

The GRAPE algorithm and its variants have become the most commonly used optimal-control methods for unitary evolutions in NMR systems \cite{Khaneja2005,Li2017}. For an $N$-qubit NMR system, the total Hamiltonian $H_{\rm{tot}}$ includes the internal term $H_{\rm{int}}$ and the radio-frequency (RF) term $H_{\rm{RF}}$:
\begin{align}
H_{\rm{tot}}=&H_{\rm{int}}+H_{\rm{RF}}, \\
H_{\rm{RF}}=&-\sum_{k=1}^{2}\gamma_{k}B_{k}[\cos(\omega^{\rm{RF}}_{k}t+\phi^{\rm{RF}}_{k})\sigma^{(1)}_{k}\notag\\
&+\sin(\omega^{\rm{RF}}_{k}t+\phi^{\rm{RF}}_{k})\sigma^{(2)}_{k}],
\end{align}
where $B_{k}$, $\omega^{\rm{RF}}_{k}$, and $\phi^{\rm{RF}}_{k}$ are the amplitude, driving frequency, and phase of the control field on the $k$th nuclear spin (with gyromagnetic ratio $\gamma_{k}$), respectively.

\begin{figure}[H]
\centering
\includegraphics[width=0.3\textwidth]{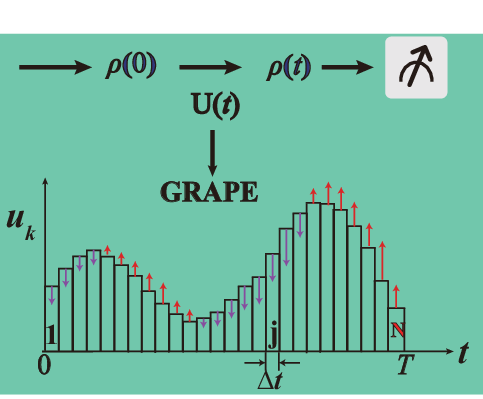}
\caption{GRAPE algorithm\cite{Wang2018,zhang2025,Khaneja2005}.}\label{fig5-SI}
\end{figure}
As shown in Fig.~\ref{fig5-SI}, the GRAPE algorithm aims to design a unitary evolution $U_{D}$ iteratively approximating the target evolution $U_{T}$, thereby determining the optimal amplitudes $B_{k}$ and phases $\phi^{\rm{RF}}_{k}$ of the control fields. The fidelity between $U_{D}$ and $U_{T}$ is defined as $F=\vert\textrm{Tr}(U_{T}^{\dagger}U_{D})\vert/2^{2}$.

Assuming the total evolution time $T$ is divided into $N$ steps ($\Delta t=T/N$) with constant control field amplitudes and phases within each step, the evolution operator for the $j$th step is:
\begin{equation}
U_{j}=e^{-i\Delta t\left[H_{\rm{int}}+\sum_{k=1}^{2}\sum_{\alpha=1}^{2}u^{(\alpha)}_{k}(j)\sigma^{(\alpha)}_{k}
\right]},
\end{equation}
where $u^{(1)}_{k}(j)=\gamma_{k}B_{k}\cos(\omega^{\rm{RF}}_{k}t_j+\phi^{\rm{RF}}_{k})$ and $u^{(2)}_{k}(j)=\gamma_{k}B_{k}\sin(\omega^{\rm{RF}}_{k}t_j+\phi^{\rm{RF}}_{k})$. The total evolution operator is $U_{D}=U_{N}U_{N-1}\dots U_{2}U_{1}$.

The gradient of the fidelity with respect to $u^{(\alpha)}_{k}(j)$ is:
\begin{align}
g^{(\alpha)}_{k}(j)&=\frac{\partial F}{\partial u^{(\alpha)}_{k}(j)}\\
&\approx -\frac{2}{2^{n}}\textrm{Re}\left[U_{T}^{\dagger}U_{N}\dots U_{j+1}(-i\Delta t\sigma^{(\alpha)}_{k})U_{j}\dots U_{1}\right].\nonumber
\end{align}
By updating $u^{(\alpha)}_{k}(j)$ as $u^{(\alpha)}_{k}(j)+\epsilon_s g^{(\alpha)}_{k}(j)$ (with $\epsilon_s$ as the iteration step) and repeating the process, the fidelity gradually increases. The algorithm terminates when the fidelity change falls below a threshold, after which measurements are performed to obtain the final result.

\end{document}